\documentclass[fleqn,twoside]{article}
\usepackage{espcrc2}

\usepackage{psfig}

\title{Indication of anomalous exchange in exclusive charmonium 
photoproduction.}

\author{A. Sibirtsev\address[i1]{Institut f\"ur Kernphysik, Forschungszentrum 
J\"ulich, D-52425 J\"ulich}%
\address[i2]{Special Research Centre for the 
Subatomic  Structure of Matter (CSSM) and Department of Physics and 
Mathematical Physics, University of Adelaide, SA 5005, Australia},
S. Krewald\addressmark[i1] and 
A.W. Thomas\addressmark[i2]}

\begin{document}

\begin{abstract}
We find that available data on $J/\Psi$ photoproduction at $|t|{\ge}$1~GeV
are energy independent over the range $22{\le}\sqrt{s}{\le}170$~GeV and 
show a very soft $t$ dependence. Through a systematic analysis of the 
data we detect a new trajectory, which manifests itself at threshold and 
is energy independent. The trajectory couples to the axial form factor 
of the nucleon and can reproduce the $J/\Psi$ exclusive photoproduction 
data at large -$t$.
\vspace{1pc}
\end{abstract}

\maketitle

Since the first measurements by the Cornell~\cite{Gittelman} and
SLAC~\cite{Camerini} Collaborations charmonium photoproduction has become an
effective tool for investigating QCD dynamics. Available
data~\cite{Gittelman,Camerini,Aubert,Frabetti,Aid,Breitweg1,Adloff,%
Breitweg2,Chekanov} are generally analysed in terms of the dependence on
the invariant collision energy squared, $s$, and four momentum 
transfer squared, $t$.

Exclusive $J/\Psi$ photoproduction can be described in terms of a $c{\bar c}$
fluctuation of the photon that interacts with the nucleon by gluon ladder
exchange. This interaction probes the gluon density, $g(x)$, and therefore
experimental results can be used to constrain the gluon parton distribution
function (PDF). The more recent DL~\cite{Donnachie6}, MRST2001~\cite{Martin} 
and CTEQ6~\cite{Pumplin} PDFs come from an analysis of recent data on $J/\Psi$
electroproduction at large $Q^2$.  Our systematic analysis~\cite{Sibirtsev1}
shows that the DL and MRST2001 PDFs reproduce available data on 
forward $J/\Psi$ photoproduction at $\sqrt{s}{>}$10~GeV.

Within the two-gluon exchange model the $s$ dependence of the forward 
$J/\Psi$ photoproduction cross section is given as
\begin{equation}
\frac{d\sigma}{dt} \propto \left[x g(x) \right]^2,
\label{twog}
\end{equation}
where $x{=}m_J^2{/}s$ and $m_J$ is the $J/\Psi$ mass. The $t$ dependence 
results from two-gluon correlation in the proton and is not 
defined by the model but can be taken~\cite{Ryskin1} to be proportional to 
the proton isoscalar EM form factor
\begin{equation}
F(t)=\frac{4m_p^2-2.8t}{4m_p^2-t} \frac{1}{(1-t/t_0)^2},
\label{emp}
\end{equation}
with $m_p$  the proton mass and $t_0$=0.71~GeV$^2$.
 
In the Regge model~\cite{Donnachie1,Donnachie2} 
the $c{\bar c}$ fluctuation interacts with the proton by exchange 
of a pomeron trajectory and
\begin{equation}
\frac{d\sigma}{dt} \propto s^{2\alpha(t)-2}\, \frac{F(t)^2}{(m_J^2-t)^2}
\,\frac{\mu^4}{(2\mu^2+m_J^2-t)^2},
\label{pom}
\end{equation}
where the last term is the form factor squared, which accounts for the
possibility that the coupling between an off-shell charm quark and the 
pomeron is not pointlike - we 
take~\cite{Donnachie1,Donnachie2,Donnachie3,Sibirtsev2}  
$\mu{=}\sqrt{1.2}$~GeV. The soft 
pomeron trajectory is $\alpha(t){=}1.08{+}0.25t$. The $s$ dependence of 
$J/\Psi$ photoproduction at high energies requires~\cite{Sibirtsev1} 
the introduction of an additional  hard
pomeron~\cite{Donnachie3} with $\alpha(t){=}1.44{+}0.1t$.

\begin{figure}[h]
\vspace*{-11mm}
\hspace*{-1mm}\psfig{file=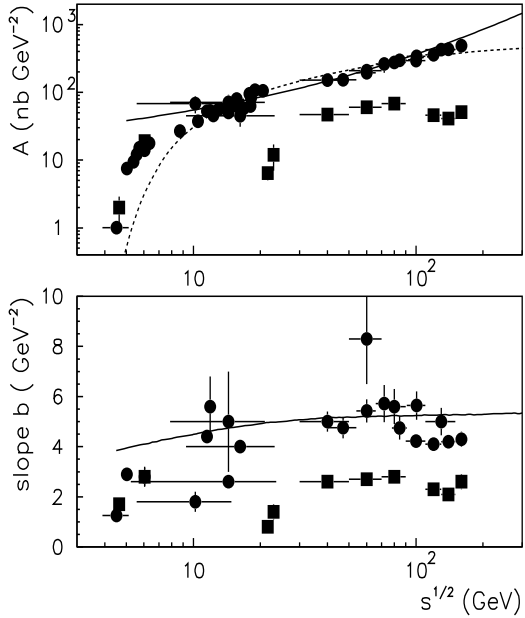,height=9.2cm,width=8.cm}
\vspace*{-16mm}
\caption[]{The forward $J/\Psi$ photoproduction cross section (upper) and 
slope $b$ of the $t$ dependence (lower) as a function of the invariant 
collision energy $\sqrt{s}$. The circles show the original
data~\cite{Gittelman,Camerini,Aubert,Frabetti,Adloff,Chekanov}, 
while the squares indicate the results of our analysis for a new, 
soft contribution. The solid lines 
are the calculations with QCD pomeron exchange, the dashed line shows 
the pQCD result from two gluon exchange with the MRST2001 PDF~\cite{Martin}.
The local slope was calculated at $t{=}$-0.5~GeV$^2$.}
\label{ajur4a}\vspace*{-2mm}
\end{figure}

The circles in Fig.\ref{ajur4a} show the results for the forward $J/\Psi$
photoproduction cross section and slope of $t$ dependence evaluated by fitting
the data as $d\sigma{/}dt{=}A\exp(bt)$. The solid line indicates a
calculation based on the Regge model with soft and hard pomeron.  The
dashed line shows the result~\cite{Sibirtsev1} from the two-gluon model with
the MRST2001 PDF. Both models reproduce the data at $\sqrt{s}{>}10$~GeV 
reasonably well and clearly indicate room for some other contribution at 
low energy.

It is important that the slope corresponding to the proton isoscalar 
EM form factor is
\begin{equation}
b= -\frac{5.6}{4m_p^2-2.8t} + \frac{2}{4m_p^2-t}
+\frac{4}{t_0-t}.
\label{slopeb1}
\end{equation}  
This depends on $t$ and can therefore be considered as a local slope.  
Furthermore $b{\simeq}4.6$ at $t{=}0$ and $b{\simeq}3.7$ at $t{=}1$~GeV$^2$.
Additional contributions from the pomeron trajectory and quark-pomeron 
form factor add to the total slope shown in Fig.\ref{ajur4a}. 
It is clear that the minimal slope is dictated by the proton EM form factor 
and the data at $\sqrt{s}{<}20$~GeV clearly indicate the need for some 
other mechanism  that does not involve the EM form factor of the proton.

\begin{figure}[h]
\vspace*{-12mm}
\hspace*{-2mm}\psfig{file=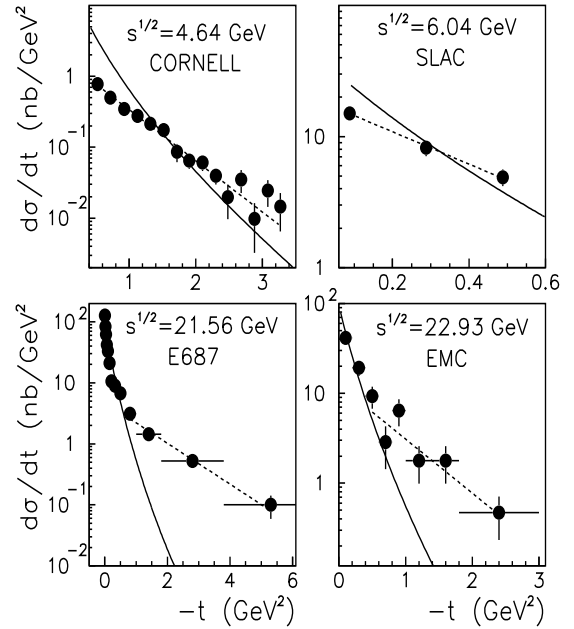,height=9.5cm,width=8.2cm}\vspace*{-13mm}
\caption[]{The $\gamma{+}p{\to}J/\Psi{+}p$ differential cross
section as a function of -$t$ measured at different $\sqrt{s}$.  The data are
from Cornell~\cite{Gittelman}, SLAC~\cite{Camerini},EMC~\cite{Aubert} and  
E687~\cite{Frabetti}.  The solid lines show the calculations
including both soft and hard pomeron exchanges. The dashed lines indicate the
fit to the soft part of the spectra.}
\vspace*{-3mm}
\label{psiju8}
\end{figure}

Low energy data~\cite{Gittelman,Camerini,Aubert,Frabetti} on $J/\Psi$
photoproduction are shown in Fig.\ref{psiju8}. The solid lines show the
calculations~\cite{Sibirtsev1,Sibirtsev2} within the  Regge model, 
while the dashed  lines illustrate the fit using an
exponential function with parameters shown by squares in Fig.\ref{ajur4a}.
The Cornell and SLAC measurements at $\sqrt{s}{<}$10~GeV are not 
reproduced by the pomeron exchange calculations.  The data collected at 
$\sqrt{s}{\simeq}$20~GeV by the EMC and E687 Collaborations are well 
described by pomeron exchange at low
$t$ and clearly indicate an additional soft contribution at $|t|{\ge}$1~GeV.

\begin{figure}[t]
\vspace*{-5mm}
\hspace*{-4mm}\psfig{file=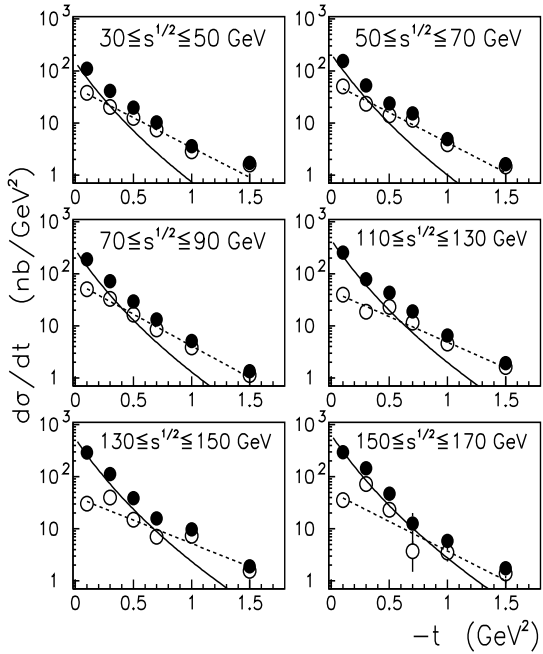,height=9.7cm,width=8.4cm}\vspace*{-13mm}
\caption[]{The solid circles are the $\gamma{+}p{\to}J/\Psi{+}p$  
differential cross section, as a function of -$t$, measured at different
invariant collision energies, $\sqrt{s}$, by the ZEUS
Collaboration~\cite{Chekanov}. The solid lines show the calculations including
both soft and hard pomeron exchanges. The open circles show the difference
between the data and these calculations. This difference has been fit by 
an exponential function shown by the dashed lines.}
\vspace*{-3mm}
\label{ajur2}
\end{figure}

Furthermore, we reanalyse the data on $J/\Psi$ photoproduction at
$30{\le}\sqrt{s}{\le}170$~GeV reported~\cite{Chekanov} recently by ZEUS
Collaboration and shown in Fig.\ref{ajur2} by solid circles in order to
extract the  soft contribution. Rather than readjusting the parameters 
of soft and hard pomeron exchanges we attribute the discrepancy between 
the calculations shown in  Fig.\ref{ajur2} by the solid lines and the 
data to an additional contribution. The open circles in  Fig.\ref{ajur2} 
show the difference between the experimental results and the calculations 
and the dashed lines show a fit to this difference using an exponential 
function with parameters shown in Fig.\ref{ajur4a} with the  squares. 

Although the above reanalysis of ZEUS data~\cite{Chekanov} cannot be 
considered definitive, because there are only two points at $|t|{\ge}$1~GeV,
it is compelling if we compare the data at
$30{\le}\sqrt{s}{\le}50$~GeV and $150{\le}\sqrt{s}{\le}170$~GeV. 
In particular, the data differ substantially  at these extreme 
energies and the calculations  reproduce the data well at higher energy 
while definitely failing to describe the 
experimental results at $30{\le}\sqrt{s}{\le}50$~GeV. But it is important that
the difference between the data and the calculations including both soft and
hard pomeron exchanges is almost the same and practically 
energy independent.

Summarizing the results of our analysis shown by squares in Fig.\ref{ajur4a},
we conclude that there is an indication for a new contribution to $J/\Psi$ 
photoproduction in addition to the to soft and hard pomeron exchanges. 
This contribution does not depend on $\sqrt{s}$. It dominates at small 
energies and can be detected at high energies at $|t|{\ge}$1~GeV$^2$ 
since it has a $t$ dependence different from that of pomeron exchange. 
Moreover, this new contribution has a slope $b{<}$3~GeV$^{-2}$, which 
means that it does not involve the proton EM form factor.
 
Within a Regge model this contribution might come from the 
exchange of a trajectory, $\alpha(t){\simeq}1.0{+}0t$, resulting 
in the  differential $J/\Psi$ photoproduction cross section given as
\begin{equation}
\frac{d\sigma}{dt} \propto s^{2\alpha(t)-2}\,
\frac{\Lambda^8}{(\Lambda^2-t)^4},
\label{axial}
\end{equation}
with cut off parameter $\Lambda{\simeq}1.2$~GeV. The choice of the form
factor is dictated by the data , which  requires a sufficiently
small local slope parameter at  $|t|{\ge}$1~GeV$^2$.

\begin{figure}[t]
\vspace*{-3mm}
\hspace*{-4mm}\psfig{file=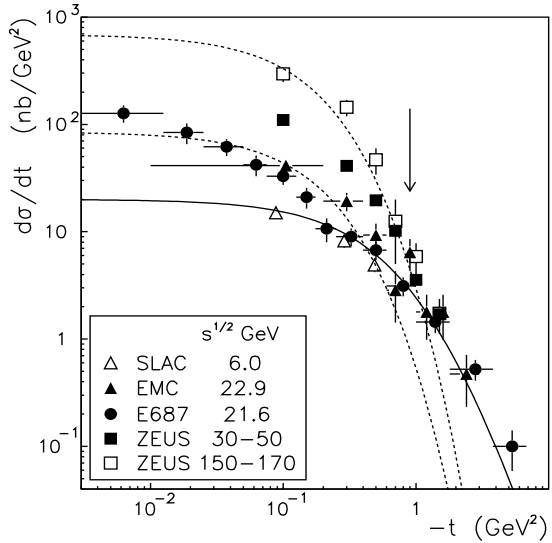,height=8.cm,width=8.4cm}\vspace*{-12mm}
\caption[]{The $\gamma{+}p{\to}J/\Psi{+}p$  
differential cross section as a function of -$t$ measured at different
energies $\sqrt{s}$. The arrow indicates the value of $t$ above which  
the data show energy independence. The dashed lines show the calculations 
with both soft and hard pomeron exchanges for $\sqrt{s}$=20~GeV and 170~GeV. 
The solid line shows the contribution from the new trajectory exchange.}
\vspace*{-4mm}
\label{ajur3c}
\end{figure}

It is clear that the available data are not good enough to draw a 
more confident conclusion and explicitly fix the parameters of the 
trajectory. A somewhat more illustrative presentation of the data is 
given in Fig.\ref{ajur3c}, where the
$\gamma{+}p{\to}J/\Psi{+}p$ differential cross section is shown as a function
of -$t$ for different energies, $\sqrt{s}$. 
Above $|t|{\simeq}$1~GeV$^2$, which is
indicated by an arrow, the data are almost energy independent.  The dashed
lines in Fig.\ref{ajur3c} show the contribution from soft and hard pomeron
exchanges at $\sqrt{s}$=20~GeV and 170~GeV. The solid line indicates the
contribution from the exchange of the new trajectory  with an absolute 
normalization $d\sigma{/}dt$=20~nb$\cdot$GeV$^{-2}$ at $t$=0. 

Our prediction is that starting from threshold the forward $J/\Psi$ 
photoproduction cross section should be of order 
20~nb$\cdot$GeV$^{-2}$ and should not be strongly energy dependent. 

Fig.\ref{ajur4a} shows that in contrast to our expectation  few 
experimental results close to threshold show energy dependence. 
The measurement at lowest energy was done by the Cornell 
Collaboration~\cite{Gittelman} with photons of energy
between 9.0 and 11.8~GeV that interacted with a beryllium target. It was
pointed out~\cite{Gittelman} that the results were not corrected for
fermi motion, which might be important since the $J/\Psi$ threshold 
corresponds to a photon energy of $\simeq$8.15~GeV. Moreover, the 
sample of $J/\Psi$ mesons was taken for the squared dielectron masses 
within the range $7.5{\div}11.0$~GeV$^2$,  that was divided into 10 
equal intervals. Since $\Gamma_{J/\Psi{\to}ee}{=}{+}5.26{\pm}0.37$~keV 
the correction from the dielectron mass resolution might be large. 

Furthermore, the SLAC measurements at a photon energies of 13 
and 15~GeV was done inclusively with a deuteron target only at one 
value of $t$ and were extrapolated to $t{=}$0 under certain assumption. 
Only the SLAC measurement at a photon  energy of 19~GeV shown in 
Fig.\ref{psiju8} was done at three different  values of $t$. 
Obviously the $J/\Psi$ photoproduction should be measured at low energies 
in order to clarify the situation.

The additional trajectory that we propose may correspond to
$f_1$ exchange with an odd signature, which
distinguishes it from pomeron exchange with even signature. We note that
unnatural parity $f_1$ exchange was proposed in Refs.~\cite{Kochelev,Oh} as an
alternative to a hard pomeron contribution~\cite{Donnachie3} in order to
describe $\rho$ photoproduction at large $|t|$. It was 
suggested~\cite{Kochelev,Oh} that $f_1$ couples to an axial form factor 
of Eq.\ref{axial} with $\Lambda{=}m_{f_1}$=1.28~GeV, which is 
slightly different~\cite{Thomas} from that extracted from muon neutrino
quasi-elastic scattering with average $\Lambda{=}1.03{\pm}0.04$~GeV. 

However the $J/\Psi$ photoproduction data discarded such an alternative. 
As is shown in Fig.\ref{ajur4a} the data on forward photoproduction 
require strong energy
dependence which is driven by hard pomeron exchange.  As well the results for
the slope at $|t|{<}$1~GeV$^2$ at $\sqrt{s}{>}$10~GeV indicate a coupling 
to the isoscalar EM form factor of the proton. The new trajectory has 
different features and should not substitute for hard pomeron exchange.

It is important that the energy independence of the data at  
large -$t$  found in our analysis also leads us to reject various 
hard processes that are expected to manifest at large -$t$ but are 
energy dependent. The new exchange can be entirely responsible for 
$J/\Psi$ photoproduction at $s{\le}10$~GeV where the pomeron exchange 
is not applicable. 

It would be of great interest to study  $J/\Psi$ 
photoproduction at low energies, which should be possible with the 
operation of  HALL~D at Jefferson Lab and to clarify whether the forward
cross section approaches a value of order 20~nb$\cdot$GeV$^{-2}$
starting from threshold. It 
will be possible and important to determine  quantum 
numbers of the new trajectory with polarization measurements.

A.S would like to acknowledge the warm hospitality and partial support 
of the CSSM during his visit. This work was supported by  the grant 
N.~447AUS113/14/0 by the Deutsche Forschungsgemeinschaft and the 
Australian Research Council.

\end{document}